# AUTOMAT PARSING OF AUDIO RECORDINGS. TESTING CHILDREN WITH DYSLALIA. -THEORETICAL BACKGROUND-

**Ovidiu-Andrei SCHIPOR[1], Titus-Marian NESTOR[2]**

*Universitatea „Ştefan cel Mare”Suceava*
*Str. Universităţii nr.9, 720225, Suceava, Romania*
[1]schipor@eed.usv.ro, [2]mnestor@stud.usv.ro

***AUTOMAT PARSING OF AUDIO RECORDINGS. TESTING CHILDREN WITH DYSLALIA. -THEORETICAL BACKGROUND-***

***Abstract*** *In this paper we present our researches regarding automat parsing of audio recordings. These recordings are obtained from children with dyslalia and are necessary for an accurate identification of speech problems. We develop the ADM algorithm and we analyze the complexity of this solution.*
*The main objectives of this task are:*

- *recording 120 children (60 with correct pronunciation and 60 with dyslalia)*
- *we must permit different audio environments during recording (some phonemes will be used for training a real time recognition system)*
- *the cost of recording devices and the children's impact must be minimized*
- *after recording is necessary to split the stream into phonemes*
- *the speech therapist's voice must be ignored*

*We utilize a digital voice recorder in High Quality mode and with VCVA (Variable Control Voice Actuator) activated. The record format is IMA-ADPCM, 16KHz and 4bits (16 bits PCM). A microphone was placed at 10 cm from mouth in order to minimize environment noise.*
*A software set of classes (C#) was created for handling audio stream (read, conversion between different format, write).*
*We also propose an original solution for placing markers in audio stream. These markers are needed for a correct parsing af full recoding.*

***Keywords*** *automat parsing, ADPCM, speech therapy, dyslalia, phoneme recording.*

## 1. Codificarea IMA-ADPCM a semnalului vocal

Pe baza tehnicii PCM (Pulse Code Modulation – Modulaţia în cod a impulsurilor) s-au dezvoltat mai multe tehnici de compresie digitală. O primă abordare presupune codificarea doar a diferenţei dintre două eşantioane consecutive şi nu a valorilor lor absolute. Această codificare se numeşte DPCM (Differential Pulse Code Modulation) şi permite o reducere la jumătate a lăţimii canalului necesar la transmisie pentru o aceeaşi calitate a semnalului.
Pentru a obţine un factor superior de compresie s-a apelat la eliminarea mai puternică a redundanţei în sensul utilizării corelaţiei dintre mai multe eşantionane. Tehnica ADPCM (Adaptatative DPCM) codifică diferenţa dintre eşantionul real (obţinut efectiv la intrare) şi un eşantion virtual, estimat pa baza eşantioanelor anterioare.
Termenul de "Adaptiv" corespunde unui calcul dinamic al funcţiei de autocorelaţie (predicţiei) într-o fereastră (pe baza unui anumit număr de eşantioane anterioare eşantionului curent). Coeficienţii funcţiei predictor se ajustează (adaptează) permanent obţinându-se un factor de compresie de 4:1 (cu păstrarea raportului semnal/zgomot). De remarcat că atunci când funcţia de predicţie se bazează doar pe eşantionul anterior şi coeficientul utilizat este 1, atunci DPCM devine un caz particular al ADPCM [1].
Există mai multe versiuni de ADPCM, integrate nativ în sistemele de operare actuale. Sub sistemul de operare Windows, fişierele ce conţin sunete în codificările PCM, DPCM, IMA-ADPCM, Microsoft-ADPCM sunt în formatul WAV. În antetul acestuia se specifică parametrii efectivi ai înregistrării (frecvenţa de eşantionare, mărimea eşantioanelor, numărul de canale, tipul codec-ului necesar) [2].
În cazul tehnicii ADPCM se utilizează doar 4 biţi pe eşantion în condiţiile unei reduceri nesemnificative a calităţii cel puţin în domeniul înregistrării de voce. Cu cât frecvenţa semnalului este mai redusă cu atât rezultatele oferite de ADPCM sunt mai bune. În cazul semnalului vocal, ponderea cea mai mare din energia codificată se află în fracţiunile de joasă frecvenţă. Aceste fracţiuni comportă deci variaţii relativ lente ceea ce permite aplicarea cu pierdere minimă de informaţie a compresiei ADPCM [3].
În ceea ce priveşte nivelul de zgomot, reducerea numărului de biţi utilizaţi determină implicit creşterea acestuia. Având în vedere însă că biţii au o pondere variabilă (în cazul unui semnal puternic au o pondere mai mare iar în cazul unui semnal mai slab au o pondere mai mică) zgomotul este direct proporţional cu nivelul semnalului util astfel încât SNR (Signal to Noise Ratio) se păstrează în limite rezonabile (sub -36dB)[4].
IMA (Interactive Multimedia Association) a dezvoltat propria schemă de codificare ADPCM considerată în general mai rapidă decât cea dezvoltată de Microsoft.

## 2. Metodologia de înregistrare

Cercetarea se referă la înregistrarea a 120 de copii (60 cu dislalie şi 60 cu pronunţie corectă). Fiecare copil a pronunţat reflectat (după logoped) 125 de foneme corespunzătoare celor 5 sunete studiate. Aparatura folosită a fost reprezentată de un reportofon digital Olympus VN-960PC în modul High Quality, cu VCVA (Variable Control Voice Actuator) activat şi senzitivitate scăzută a microfonului. Formatul de înregistrare este IMA – ADPCM, un singur canal, cu frecvenţa de eşantionare de 16kHz (eşantioane pe secundă) şi cuantizarea pe 4 biţi (echivalent cu 16 biţi PCM)[6].
Înregistrarea nu s-a realizat într-un spaţiu izolat fonic deoarece metodologia pusă la punct în cadrul acestei cercetări trebuie să fie suficient de robustă pentru a fi aplicată în condiţii obişnuite (domiciliul logopatului). Pentru a reduce influenţa zgomotului existent în încăpere s-a utilizat un microfon de dimensiuni reduse amplasat pe gulerul copilului, centrat în raport cu zona frontală, conectat la intrarea corespunzătoare a reportofonului. Amplasarea la o distanţă foarte redusă (aproximativ 10 cm de cavitatea bucală) a necesitat stabilirea la nivel scăzut a senzitivităţii microfonului.
Pentru a permite înregistrarea doar a fonemelor spuse de copil, s-a recurs la o extensie a reportofonului digital în sensul conectării unui întrerupător normal închis în paralel cu microfonul. În mod normal (**starea 1 – S1**) întrerupătorul nu este apăsat, microfonul este scurtcircuitat şi nu transmite nimic reportofonului (exact ca în condiţiile unei linişti depline). După aproximativ o secundă, intră în funcţiune sistemul VCVA care opreşte efectiv şi înregistrarea (**starea 2 – S2**). În momentul în care se doreşte înregistrarea logopatului se acţionează şi se ţine apăsat întrerupătorul. Aceasta determină deschiderea contactului, apariţia unei impedanţe infinite în paralel cu microfonul şi deci direcţionarea semnalului de la microfon către reportofon (**starea 3 – S3**). Imediat după ce fonemul a fost pronunţat, logopedul eliberează întrerupătorul şi se revine la starea iniţială.
În acest fel, înregistrarea va conţine exclusiv fonemele dorite a fi înregistrare separate de zone de „linişte" în general mai mari de o secundă. Schematic, stările pot fi modelate cu ajutorul următorului graf.

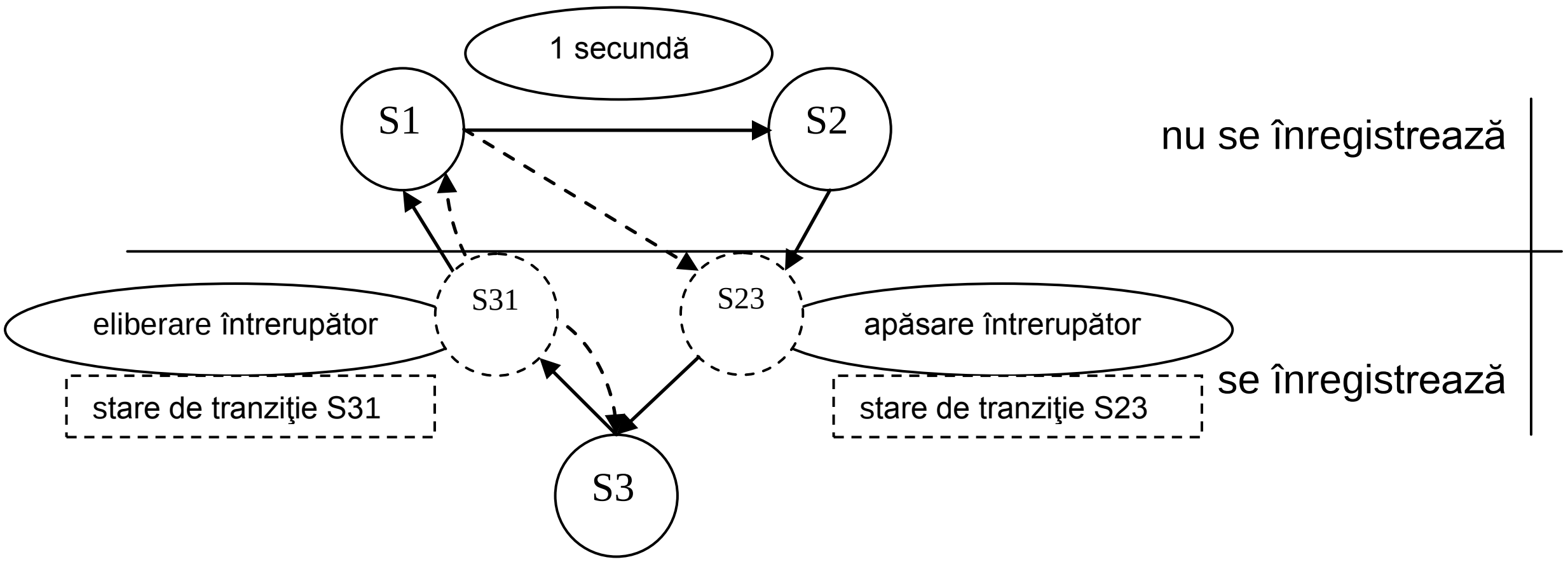

**Figura 1. Graful stărilor în cazul înregistrării fonemelor**

Aşa cum este prezentat şi în figură, apar două stări intermediare (stările **S23** şi **S31**) aflate tot în domeniul „vizibil" al înregistrării. Aceste stări tranzitorii, cu durata de aproximativ 13 milisecunde, au un aspect specific relativ uşor identificabil în stream-ul de analizat. Ele îndeplinesc astfel rolul de veritabili marcatori (delimitatori, separatori) ai fonemelor înregistrate.

În graful prezentat se pot identifica încă două drumuri (marcate punctat) mai puţin importante în procesul de analiză a stream-ului audio:

- drumul **S1-S23** - dacă acţionarea butonului se realizează înainte de intrarea în regim VCVA;
- drumul **S3-S1-S3** (fără trecere prin **S31**) - dacă intrarea în mod VCVA se realizează ca urmare a pauzei în pronunţie a copilului şi nu ca urmare a eliberării întrerupătorului.

## 3. Parametrii specifici ai marcatorilor

Marcatorii corespunzători stărilor de tranziţie **S23** şi **S31** au forme specifice ceea ce permite identificarea lor şi implicit separarea fonemelor. Aceste stări de tranziţie apar ca urmare a răspunsului sistemului la un semnal tip treaptă [7]. Figura următoare prezintă imaginea unui asemenea marcator.Se identifică trei faze, primele două de amplitudini mari şi separate printr-o schimbare bruscă de semn şi ultima caracterizată printr-o scădere logaritmică spre valoarea 0.
Parametrii specifici ai unui marcator sunt:

- $\mathbf{A_1}$ – amplitudinea medie a semnalului în prima fază a tranziţiei (0.8*max ≈ 26.000);
- $\mathbf{A_2}$ – amplitudinea medie a semnalului în a doua fază a tranziţiei (0,7*max ≈ 23.000);
- $\mathbf{T_1}$, $\mathbf{T_2}$ – timpii cât durează acţiunea primele două faze (40 eşantioane ≈ 3ms);

- **Tr** – timpul de reglare (durata stării de tranziţie) considerat până la intrarea amplitudinii în intervalul +/- 5% (140 eşantioane ≈ 13ms).

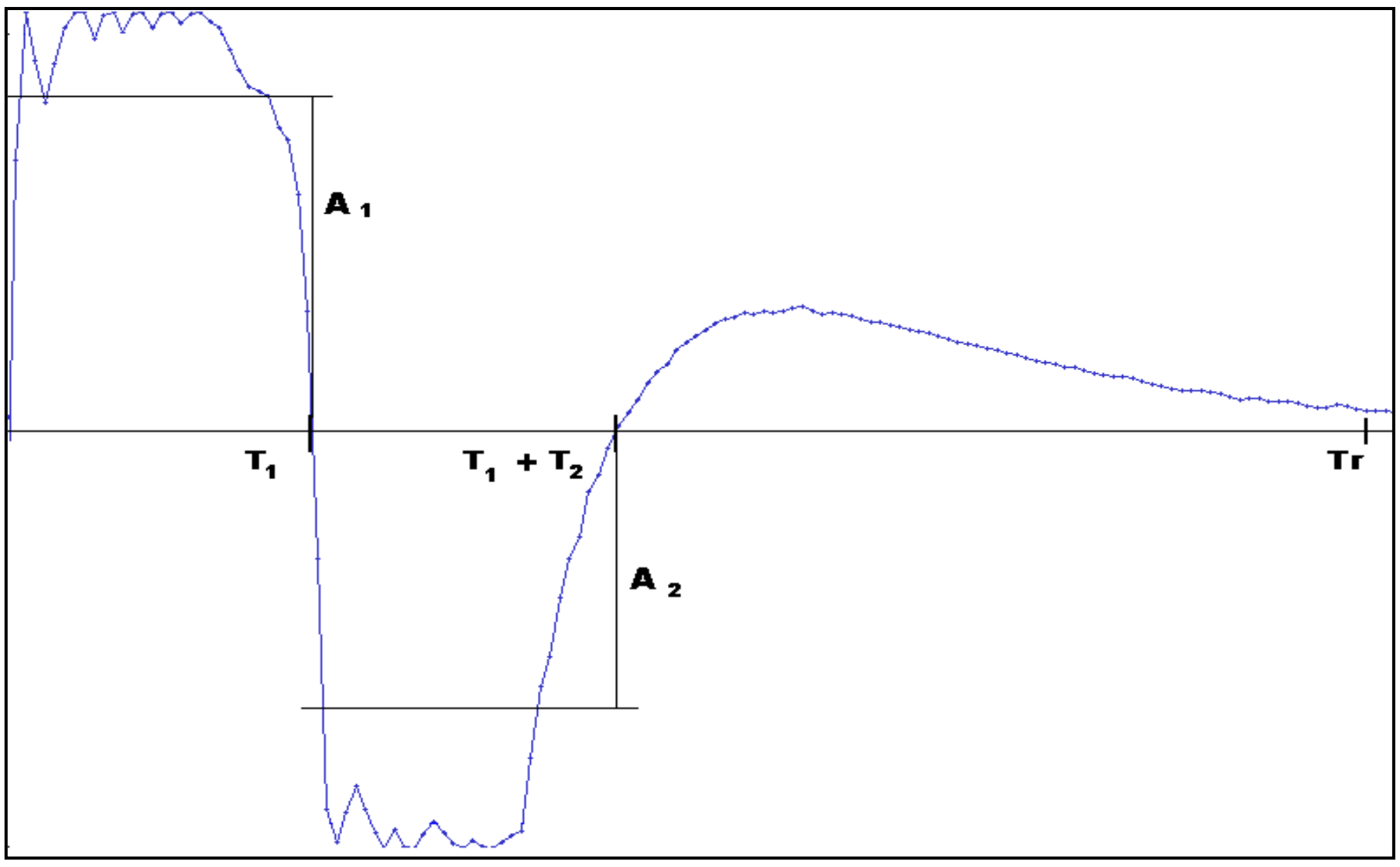

**Figura 2. Stare de tranziţie (marcator)**

Valorile precizate sunt obţinute în urma analizei statistice a peste 1000 de marcatori.
Fiecare fonem este separat de următorul printr-o secvenţă **S31-S1-(S2)-S23** adică printr-o secvenţă **marcator – linişte - marcator**. Astfel, separarea fonemelor poate fi realizată fără probleme de îndată ce se pune la punct un algoritm de detecţie a marcatorilor.

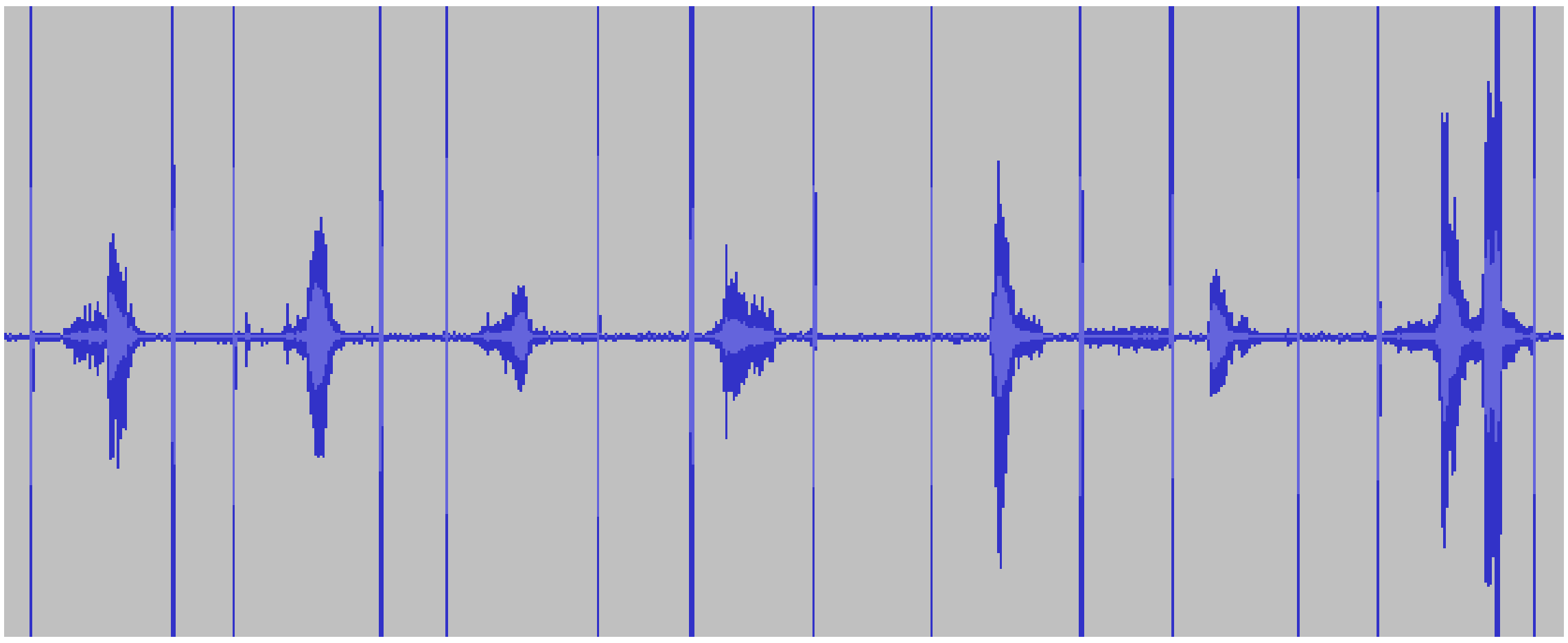

**Figura 3. Foneme separate prin marcatori**

## 4. Algoritm de detecţie a marcatorilor

Algoritmul de detecţie a marcatorilor notat în continuare **ADM** îşi propune obţinerea unui vector **M** care să conţină poziţiile de început ale marcatorilor dintr-un stream audio reprezentat printr-un vector **ST** (a cărui lungimea este numărul de eşantioane).
Structura algoritmului ADM este relativ simplă şi se bazează pe valorile foarte mari (în valoare absolută) ale eşantioanelor în cadrul primelor două etape de tranziţie precum şi pe schimbarea de semn a valorii semnalului. Reprezentarea este realizată în pseudocod [8].

Se consideră următoarele variabile globale:

```
A - valoarea medie a esantioanelor in cadrul etapelor 1 si 2
T - timpul minim cat trebuie sa dureze cele doua etape pentru a
    considera zona marcator
P - procentul de esantioane peste A pentru a considera zona marcator
Tr - timpul de reglare

ADM (ST,M)
       int i,j=0
       pentru i=0,||ST||-T,1 executa
             daca |ST[i]|>A atunci
                   daca esteMarcator(ST,i) atunci
                         M[j++]=i
                         i+=Tr
@      @     @     @

esteMarcator (ST,i)
       int N=0,nr_semn=0
       int semn = ST[i]>0 ? 1 : -1
       pentru i=1,T,1 executa
             daca |ST[i]|>A atunci
                   N++
             @
             daca semn*ST[i] < 0 atunci
                   nr_semn++
             @
       @
       daca N*100/T >= P SI nr_semn<=1 atunci
             return 1
       altfel
             return 0
@      @
```

Se apelează procedura esteMarcator doar în cazul apariţiei unui eşantion promiţător (cu valoare superioară referinţei V considerate). După identificarea unui marcator, căutarea următoarei ocurenţe se realizează după numărul Tr de eşantioane (timpul mediu de reglare adică timpul cât durează un marcator complet).

Procedura esteMarcator verifică doar primele T eşantioane din zona candidat pentru marcator. Această abordare îmbunătăţeşte simţitor performanţa algoritmului (T reprezintă doar 40% din lungimea întregului marcator) fără a prejudicia procentul de reuşită. Condiţia necesară şi suficientă pentru ca o zonă de lungime T să fie începutul unui marcator este ca ea să conţină o singură schimbare de semn şi mai mult de P% eşantioane mai mari în valoare absolută decât A.

Rezultate optime (identificare peste 97%) s-au obţinut cu **P**=75, **Tr**=350, **T**=50 eşantioane şi **A**=27000. Un factor decisiv îl reprezintă şi nivelul sonor al înregistrării care trebuie să se încadreze în limitele +/- 0.8*max.

## 5. Analiza complexităţii şi performanţelor algoritmului în funcţie de valorile parametrilor

În continuare se utilizează următoarele notaţii:

**s**=||S||

**m**=||**M**||

**c** =||**C**|| – candidaţii la început de marcator din S

**np = P*100/T** – numărul de eşantioane mai mari ca V pentru ca o zonă de lungime T să fie marcator

Numărul de operaţii elementare ale algoritmului ADM variază în jurul valorii s care va fi considerată valoare de referinţă. Pentru a obţine expresia matematică a numărului de operaţii elementare în cazul cel mai favorabil / defavorabil se porneşte de la constatarea că un candidat acceptat determină o economie de Tr-T operaţii elementare în algoritmul ADM în timp ce un candidat neacceptat determină un exces de T operaţii elementare.
Cazul cel mai favorabil corespunde astfel situaţiei în care toţi candidaţii evaluaţi sunt acceptaţi adică m==c. Se obţine expresia:

$$\mathbf{T_{bestADM} = s - m\,(Tr\text{-}T)} \qquad \textbf{[1]}$$

Experimental se constată că există o relaţie de directă proporţionalitate între s şi m în sensul că m este o fracţiune a din s. Se deduce astfel relaţia:

$$\mathbf{T_{bestADM} = s\,[1 - a\,(Tr\text{-}T)]} \qquad \textbf{[2]}$$

Cazul cel mai defavorabil corespunde situaţiei în care toţi candidaţii evaluaţi sunt respinşi adică m==0. Se obţine expresia:

$$\mathbf{T_{worstADM} = s * c * T} \qquad \textbf{[3]}$$

Numărul de candidaţi este şi el în relaţie de proporţionalitate directă cu s în sensul că reprezintă o fracţiune b din s. Se deduce astfel relaţia:

$$\mathbf{T_{worstADM} = s^2 * b * T} \qquad \textbf{[4]}$$

Trebuie de precizat că ambii coeficienţi a şi b sunt puternic subunitari astfel încât, complexitatea algoritmului ADM este:

$$\mathbf{\mathcal{C}_{ADM} = O(n)} \qquad \textbf{[5]}$$

De asemenea, o condiţie esenţială pentru reducerea candidaţilor neacceptaţi este înregistrarea eşantioanelor vocale într-un domeniu de valori situat între +/- 0.8 * max.
Alegerea parametrilor T,V şi P influenţează puternic procentul de identificare corectă. Un T prea mare determină neacceptarea unor candidaţi care în realitate sunt marcatori. Pe de altă parte, o valoare prea mică poate determina acceptarea unor candidaţi falşi (vârfuri accidentale de semnal). Parametrii V şi P prezintă o discuţie similară.
Experimental, s-a constatat un optim de recunoaştere (peste 97%) cu următoarele valori:

**P=75, T=50, A=27000**

La modificarea condiţiilor de înregistrare (schimbarea reportofonului, microfonului sau a întrerupătorului) nu este de regula necesarea re-ajustarea acestor parametrii.

**References:**

[1] – Holzinger Andreas (2005) - *Multimedia Basics Technology*, Editura Frewall Media.
[2] – *www.microsoft.com.*
[3] – Coulter Doug (2000) – *Digital Audio Processing*, Editura CMP Books.
[4] – Clarkson Peter (2001) – *Optimal and Adaptative Signal Processing, third edition*, CRC Press.
[5] – *www.wikipedia.org.*
[6] – *www.olympus.com.*
[7] – Diatcu E. (2000) – *Elemente fundamentale ale teoriei sistemelor şi calculatoarelor*, Editura Hiperion XXI, Bucureşti.
[8] – Pentiuc Ştefan-Gheorghe – *Structuri de date şi algoritmi fundamentali*, Editura Universităţii "Ştefan cel Mare", Suceava